\documentclass[10pt]{iopart}
\usepackage{iopams}  

\begin{document}

\title{Non-linear generalized elasticity of icosahedral quasicrystals}

\author{M Ricker\footnote[3]{To whom correspondence should be
    addressed (mricker@itap.physik.uni-stuttgart.de)} and H-R Trebin}

\address{Institut f\"ur Theoretische und Angewandte Physik,
  Universit\"at Stuttgart, Pfaffenwaldring 57, 70550 Stuttgart,
  Germany}

\begin{abstract} Quasicrystals can carry, in addition to the
  classical phonon displacement field, a phason displacement field,
  which requires a generalized theory of elasticity. In this paper,
  the third-order strain invariants (including phason strain)
  of icosahedral quasicrystals are determined.
  They are connected with 20 independent third-order elastic
  constants. By means of non-linear elasticity, phason strains with
  icosahedral irreducible $\rm{\Gamma}^{4}$-symmetry can be obtained
  by phonon stress, which is impossible in linear elasticity.

\end{abstract}


\section{Introduction}

Apart from the ordinary phonon degrees of freedom, quasicrystals (QC) have 
phason degrees of freedom, referring to relative shifts of the constituent
density waves \cite{J1,J2}. Therefore, and due to their lack of
translational order, quasicrystalline structures are usually
constructed as irrational cut of a decorated hyperspace structure
by physical space $E^{\parallel}$ \cite{J3,J4}. The phason degrees
of freedom are connected with the displacement field along the
orthogonal space $E^{\perp}$. The generalized elasticity is described
in terms of spatially varying phonon and phason displacement
fields \cite{J1,J2}.

Icosahedral QC have three phonon and three phason degrees of
freedom and associated components of a displacement field. Linear
elastic theory provides five independent second-order elastic
constants, two belonging to pure phonon elasticity, two to pure phason
elasticity and one to a coupling between phonons and phasons.

Within linear phonon elasticity, icosahedral QC behave essentially
like isotropic media \cite{J5}. Faithful icosahedral symmetry exists
for physical properties described by tensors of rank $N \ge 5$ only
\cite{J6}.  Accordingly, the non-linear elasticity of icosahedral QC is
anisotropic \cite{J7}.

Fundamental research on classical non-linear elasticity has been
performed many years ago \cite{J8,J9}. The authors of
\cite{J6,J10,J11,J12,J13} 
have already determined the four linearly independent, icosahedral
elastic tensors of rank six, related to third-order phonon elastic
invariants. Contrary to this, in the isotropic case one has only three
independent third-order phonon elastic invariants, or elastic 
constants. Ishii \cite{J14} has calculated the pure phason third-order
icosahedral invariants. The aim of this paper is to generalize the classical
non-linear elasticity and to determine all third-order icosahedral
elastic invariants, which occur when phason strains are
included. The idea leading to this work was to find a possibility to
generate phason $\Gamma^{4}$-strain in icosahedral QC by phonon
stress, which is impossible within linear elasticity.

\section{Generalized elastic theory of icosahedral QC}

According to their icosahedral diffraction pattern, the mass density
of icosahedral QC is a sum of density waves indexed by a reciprocal
lattice $L$ of icosahedral symmetry: $ \rho(\bi{x}) = \sum_{\bi{k} \in
L} \rho_{\bi{k}} \exp(i\bi{k}\cdot\bi{x}) = \sum_{\bi{k} \in L}
|\rho_{\bi{k}}| \exp[i(\bi{k}\cdot\bi{x} + \phi_{\bi{k}})]$. The
phases $\phi_{\bf{k}}$ of the basis vectors of $L$ are six degrees of
freedom \cite{J2}, parametrized by the phonon and phason
displacement fields $\bi{u}$ and $\bi{w}$ via $ \phi_{\bi{k}} =
\phi_{\bi{k},0} - \bi{k}^{\parallel} \cdot \bi{u} - \bi{k}^{\perp}
\cdot \bi{w} \, . $ Here, $\bi{k}^{\parallel}=\bi{k}$ and
$\bi{k}^{\perp}$ are the projections of reciprocal six-dimensional
hyperlattice vectors onto $E^{\parallel}$ and $E^{\perp}$,
respectively. The phonon and phason displacement fields $\bi{u}$ and
$\bi{w}$ are the projections of the hyperspace displacement
field $\bi{u} \oplus \bi{w}$ onto $E^{\parallel}$ and
$E^{\perp}$, respectively. 

We denote the position of a point in the undistorted QC $\bi{a}$ and the
corresponding position in the distorted structure $\bi{x}$,  where $\bi{x} =
\bi{u} + \bi{a}$. In the Lagrangian scheme, all quantities depend on the
variable $\bi{a}$ \cite{J8}. The phonon strain tensor $\bfeta^{u}$ has its
components of the classical symmetric form
\begin{equation}
\label{eq1}
\eta^{u}_{ij} = \frac{1}{2} \Big( \frac{\partial u_{i}}{\partial a_{j}} + 
\frac{\partial u_{j}}{\partial a_{i}} \Big) + \frac{1}{2} \, \frac{\partial
  u_{k}}{\partial a_{i}} \frac{\partial u_{k}}{\partial a_{j}}
\end{equation}
or, written in terms of the Jacobian $F_{ij} = \frac{\partial x_{i}}{\partial
a_{j}}$, $\eta^{u}_{ij} = \frac{1}{2} (F_{ki} \, F_{kj} - \delta_{ij})\,$. This
strain tensor is free of rigid rotations. The phason displacement gradient
$\frac{\partial \bi{w}}{\partial \bi{a}}$ splits into a $\Gamma^{4}$ and a 
$\Gamma^{5}$ part (see (\ref{eq9})), and both are assumed to increase the elastic
energy \cite{J2,J15}. Therefore, we have a phason strain tensor $\bfeta^{w}$ with
\begin{equation}
\label{eq2}
\eta^{w}_{ij} =  \frac{\partial w_{i}}{\partial a_{j}} \, .
\end{equation}
In the linear limit $|\frac{\partial u_{i}}{\partial a_{j}}|
\ll 1$, the components of $\bfeta^{u}$ take their 
well-known shape $\eta^{u}_{ij} = \frac{1}{2} 
( \frac{\partial u_{i}}{\partial a_{j}} + \frac{\partial u_{j}}
{\partial a_{i}})$.  
   
The isothermal Helmholtz free energy $F(\bfeta^{u},\bfeta^{w})$ per
undistorted volume can be expanded into the Taylor series
\begin{equation}
\label{eq3}
\fl F = \frac{1}{2} \, C_{ijkl}^{ab} \, \eta^{a}_{ij} \, 
\eta^{b}_{kl} + \frac{1}{6} \, C_{ijklmn}^{abc} \, 
\eta_{ij}^{a} \, \eta_{kl}^{b} \,
\eta_{mn}^{c} + \ldots = \frac{1}{2} \, C_{ij}^{ab} \, \eta^{a}_{i} \, 
\eta^{b}_{j} + \frac{1}{6} \, C_{ijk}^{abc} \, 
\eta_{i}^{a} \, \eta_{j}^{b} \,
\eta_{k}^{c} + \ldots ,
\end{equation}
with $C_{ijkl}^{ab}$ being second-order and $C_{ijklmn}^{abc}$ third-order
cartesian elastic constants due to Brugger \cite{J16}, which is perhaps 
the most familiar notation ($i,j,k,l,m,n \in \{1,2,3\}$;  $a,b \in \{u,w\}$). 
In the right part of (\ref{eq3}), the irreducible strain components of \ref{A1} are used.
Here we have, e.g., $i \in \{1, \ldots, 6\}$ if $a = u$ and $i \in \{1, \ldots, 9\}$
if $a = w$. Because of the index permutation symmetries and the symmetries of the QC,
not all of these $C'$s are independent. If the elastic energy is to be written
with independent elastic constants $C^{ab}_{i}$ and  $C^{abc}_{i}$ of
second and third order only, one has to use the invariants $I^{ab}_{i}$ and 
$I^{abc}_{i}$ of the generalized elasticity:
\begin{equation}
\label{eq4}
F = C_{i}^{ab} I^{ab}_{i} \, + C_{i}^{abc} I^{abc}_{i} + \ldots \, .
\end{equation}
These invariants (and also the expansions (\ref{eq3})) must fulfill the condition
$I(g\bfeta^{u},g\bfeta^{w}) = I(\bfeta^{u},\bfeta^{w})$ for any transformation
$g$ of the icosahedral group $Y$ (or $Y_{h}$). Clearly, symmetries like
$I^{uuw}_{i} =  I^{uwu}_{i} = \ldots \,$ and
$C^{uuw}_{i} =  C^{uwu}_{i} = \ldots \,$ are assumed to hold.
The elastic constants of (\ref{eq3}) follow from appropriate repeated 
differentiation with respect to components of $\bfeta^{u}$ and 
$\bfeta^{w}$. 

A generalization of the classical result \cite{J16} shows that the generalized
Piola-Kirchhoff stresses $\bi{t}^{u}$ and $\bi{t}^{w}$, which are measured in
the undistorted state, have components
\begin{equation}
\label{eq5}
\fl t^{a}_{ij} = \frac{\partial F}{\partial \eta^{a}_{ij}} = C^{ab}_{ijkl}
\eta^{b}_{kl} + \frac{1}{2} \, C^{abc}_{ijklmn}
\eta^{b}_{kl} \, \eta^{c}_{mn} + \ldots = C_{k}^{ab} 
\frac{\partial I^{ab}_{k}}{\partial \eta^{a}_{ij}}
+ C_{k}^{abc} \frac{\partial I^{abc}_{k}}{\partial
  \eta^{a}_{ij}} 
+ \ldots \, . 
\end{equation}
The irreducible components of $\bi{t}^{u}$ and $\bi{t}^{w}$ have the same form
as those of $\bfeta^{u}$ and $\bfeta^{w}$, and it is $t^{a}_{i} = 
\frac{\partial F}{\partial \eta^{a}_{i}}$, where the possible pairs of $(a,i)$
are again evident from \ref{A1}. Cauchy stresses $\bsigma^{u}$ and
$\bsigma^{w}$ are measured in the distorted state. They follow immediately
from the Piola-Kirchhoff stresses \cite{J8}. Because $E^{\perp}$ remains
unchanged even for a finite deformation, the phasonic case must be 
treated with some care: 
\begin{equation}
\label{eq6}
\sigma^{u}_{ij} = \Delta^{-1} \, F_{ik} \, F_{jl} \, t^{u}_{kl}  \, , \\
\sigma^{w}_{ij} = \Delta^{-1} \, F_{jk} \, t^{w}_{ik} \, ,
\end{equation}
where $\Delta = \mathrm{det} \bi{F}$.

Since our intention is to produce certain strains by means of applied 
stresses, we should rather work with the isothermal Gibbs enthalpy
$G(\bi{t}^{u},\bi{t}^{w})$ \cite{J16} 
\begin{equation}
\label{eq7}
G = S_{i}^{ab} I^{ab}_{i} \, + S_{i}^{abc} I^{abc}_{i} + \ldots \, .
\end{equation}
$S^{ab}_{i}$ and $S_{i}^{abc}$ are independent elastic
compliances, and the same invariants as in (\ref{eq4}) appear, but this time
formulated with components of $\bi{t}^{u}$ and $\bi{t}^{w}$. One obtains
the strain-stress relations
\begin{equation}
\label{eq8}
\fl \eta^{a}_{ij} = - \frac{\partial G}{\partial t^{a}_{ij}} = -S^{ab}_{ijkl}
t^{b}_{kl} - \frac{1}{2} \, S^{abc}_{ijklmn}
t^{b}_{kl} \, t^{c}_{mn} + \ldots = - S_{k}^{ab} 
\frac{\partial I^{ab}_{k}}{\partial t^{a}_{ij}}
- S_{k}^{abc} \frac{\partial I^{abc}_{k}}{\partial
  t^{a}_{ij}} - \ldots \, . 
\end{equation}
The irreducible form hereof is obvious.

\section{The elastic invariants} 
\label{sec3}

First thing of all is to note the transformation behaviour of the strain 
tensors \cite{J2}, which follows from the transformation behaviour of 
vectors in $E^{\parallel}$ and $E^{\perp}$:
\begin{equation}
\label{eq9}
u: \;\; (\Gamma^{3} \otimes \Gamma^{3})_{s} = \Gamma^{1} \oplus
\Gamma^{5} \, , \\
w: \;\; \Gamma^{3'} \otimes \Gamma^{3} = \Gamma^{4} \oplus
\Gamma^{5} \, .
\end{equation}
Index $s$ means symmetrized. The irreducible components of the strain 
tensors are given in \ref{A1}, and the associated transformation matrices 
are displayed in Table \ref{tabtwo}. They are deduced from the transformation
of the cartesian strain tensors
$\bfeta^{a} = \eta^{a}_{ij} \, \bfeta^{a}_{ij}$, where $\bfeta^{a}_{ij}$
is a basis deformation with component $ij$ being 1 and all others 0:
$g\bfeta^{a} = \eta^{a}_{ij} \, g\bfeta^{a}_{ij} =
(g\bfeta^{a})_{ij} \, \bfeta^{a}_{ij}\,$, where
$(g\bfeta^{u})_{ij}=D^{3}_{ik}(g) \, D^{3}_{jl}(g) \, \eta^{u}_{kl}\,$,
$(g\bfeta^{w})_{ij}=D^{3'}_{ik}(g) \, D^{3}_{jl}(g) \,\eta^{w}_{kl}\,$, and
$\bi{D}^{3}(g)\,$, $\bi{D}^{3'}(g)$ are the coordinate transformation
matrices of Table \ref{tabtwo}. 

The characters of symmetrized product representations $D =
(\tilde{D}\otimes \tilde{D})_{s}$ and $D = (\tilde{D}\otimes
\tilde{D} \otimes \tilde{D})_{s}$, respectively, of one and the same
representation $\tilde{D}$, which the terms occuring in (\ref{eq3}) 
transform after, are
\begin{equation}
\label{eq10}
\eqalign{ \chi^{D}(g) = \frac{1}{2}\big[ \, \chi^{\tilde{D}}(g)^{2}
+\chi^{\tilde{D}}(g^{2})\big] \quad \mbox{and} \\ \chi^{D}(g) =
\frac{1}{6}\big[ \, \chi^{\tilde{D}}(g)^{3} + 3 \, \chi^{\tilde{D}}
(g^{2}) \chi^{\tilde{D}} (g) + 2 \, \chi^{\tilde{D}}(g^{3})\big] \, .}
\end{equation}
Herewith, the following Clebsch-Gordan series for the representations acting
in the respective symmetrized product spaces of second-order tensors are 
straightforward and given below in the order 
$uu$, $uw$, $ww$, $uuu$, $uuw$, $uww$, $www$ \cite{J17}:
\begin{equation}
\label{eq11}
\eqalign{
\fl \, [(\Gamma^{1}\, \oplus \,\Gamma^{5})\,\otimes \,(\Gamma^{1}\,
\oplus \, \Gamma^{5})]_{s}&  \, = \,2\, \Gamma^{1}\, \oplus\, 
\Gamma^{4} \,\oplus \,3\, \Gamma^{5}\, ,
\\
\fl \, (\Gamma^{1}\, \oplus \,\Gamma^{5})\, \otimes \,(\Gamma^{4}\, \oplus
\, \Gamma^{5})& \, = \, \Gamma^{1}\, \oplus \, 2
\,\Gamma^{3}\,\oplus\, 2\,\Gamma^{3'}\, \oplus \,4\,\Gamma^{4}
\, \oplus \, 5 \, \Gamma^{5} \, ,
\\
\fl \, [(\Gamma^{4}\, \oplus \,\Gamma^{5})\, \otimes \,(\Gamma^{4}\, \oplus
\, \Gamma^{5})]_{s}& \, = \, 2\, \Gamma^{1}\, \oplus
\,\Gamma^{3}\,\oplus\, \Gamma^{3'}\, \oplus \,3\,\Gamma^{4}
\, \oplus \, 5 \, \Gamma^{5} \, , 
\\
\fl \, [(\Gamma^{1} \oplus \Gamma^{5})\otimes
(\Gamma^{1} \oplus \Gamma^{5})\otimes(\Gamma^{1}
 \oplus \Gamma^{5})]_{s}
&\, = \,4\,\Gamma^{1} \oplus \Gamma^{3} \oplus 
\Gamma^{3'} \oplus 4\,\Gamma^{4} \oplus 6\,\Gamma^{5} \,,
\\
\fl \, [(\Gamma^{1} \oplus \Gamma^{5})\otimes(\Gamma^{1}
 \oplus \Gamma^{5})]_{s}\otimes(\Gamma^{4} \oplus 
\Gamma^{5})
&\, = \,4\,\Gamma^{1} \oplus 8\,\Gamma^{3} \oplus 
8\,\Gamma^{3'} \oplus 13\,\Gamma^{4} \oplus 17
\,\Gamma^{5}\,,
\\
\fl \, (\Gamma^{1} \oplus \Gamma^{5})\otimes
[(\Gamma^{4} \oplus \Gamma^{5})\otimes(\Gamma^{4}
 \oplus \Gamma^{5})]_{s}
&\, = \,7\,\Gamma^{1} \oplus 11\,\Gamma^{3} \oplus 11
\,\Gamma^{3'} \oplus 18\,\Gamma^{4} \oplus 25\,\Gamma^{5}\,,
\\
\fl \, [(\Gamma^{4} \oplus \Gamma^{5})\otimes
(\Gamma^{4} \oplus \Gamma^{5})\otimes(\Gamma^{4} \oplus 
\Gamma^{5})]_{s}
&\, = \,5\,\Gamma^{1} \oplus 7\,\Gamma^{3} \oplus 7\,
\Gamma^{3'} \oplus 12\,\Gamma^{4} \oplus 14\,\Gamma^{5}\,.
}
\end{equation}
We see that we have 
the following numbers of invariants: 2 ($uu$), 1 ($uw$), 2 ($ww$), 
4 ($uuu$), 4 ($uuw$), 7 ($uww$), 5 ($www$). These numbers have already
been calculated in an earlier work \cite{J18}.

The second-order invariants can readily be written as simple scalar products
 $\bfeta^{u,1} \cdot \bfeta^{u,1}$, $\bfeta^{u,5} \cdot \bfeta^{u,5}$, 
$\bfeta^{u,5} \cdot \bfeta^{w,5}$, $\bfeta^{w,4} \cdot \bfeta^{w,4}$ and 
$\bfeta^{w,5} \cdot \bfeta^{w,5}$  of vectors containing the irreducible 
strain components (see \ref{A1} and  \cite{J19}). 
Of course, most of the third-order invariants are more complicated.

In Table \ref{tabone}, the symmetrized third-order vector spaces are
specified in more detail. Some terms must be weighted with factors to
become components for an orthonormal basis set and to transform 
orthogonal among all others.
\begin{table}
\caption{Tabulation of symmetrized third-order vector spaces. 
In the last column, all possible third-order expressions
consisting of irreducible strains, referring to
orthonormal basis sets, are listed. \label{tabone}}  
\begin{indented}
  \lineup
\item[]\begin{tabular}{@{}*{7}{l}}
\br                              
Case & Dimension $dim$ of & Components with respect & \\
&  symmetrized space & to an orthonormal basis set & \cr 
\mr
$uuu$ & 56  & 
$(\eta^{u}_{i})^{3}$, $\sqrt{3} \, (\eta^{u}_{i})^{2} \eta^{u}_{j}$, 
$\sqrt{6} \, \eta^{u}_{i} \eta^{u}_{j} \eta^{u}_{k}$ & ($i \neq j \neq k \neq
i$) \cr \ms
$uuw$ & 189 &
$(\eta^{u}_{i})^{2} \eta^{w}_{k}$, 
$\sqrt{2} \, \eta^{u}_{i} \eta^{u}_{j} \eta^{w}_{k}$ & ($i \neq j$) \cr \ms
$uww$ & 270 &
$\eta^{u}_{i} (\eta^{w}_{j})^{2}$, 
$\sqrt{2} \, \eta^{u}_{i} \eta^{w}_{j} \eta^{w}_{k}$ & ($j \neq k$) \cr \ms
$www$ & 165 &
$(\eta^{w}_{i})^{3}$, $\sqrt{3} \, (\eta^{w}_{i})^{2} \eta^{w}_{j}$, 
$\sqrt{6} \, \eta^{w}_{i} \eta^{w}_{j} \eta^{w}_{k}$ & ($i \neq j \neq k \neq
i$) \cr 
\br
\end{tabular}
\end{indented}
\end{table}
The third-order elastic invariants are the components for the basis vectors of
the identity representation.

Basis vectors for an irreducible group representation $\alpha$ are obtained by
means of the projection operators $\bi{P}^{\alpha}_{lk} =
\frac{d_{\alpha}}{|G|} \sum_{g
\in G} D^{\alpha \,*}_{lk} (g) \bi{D}(g)$ \cite{J20}, with
$\bi{D}(g)$ being the linear operators acting on the full (reducible) 
vector space (see Table \ref{tabone}). These
projectors have the property $ \bi{P}^{\alpha}_{lk} \bi{e}^{\beta}_{ij} =
\delta_{\alpha \beta} \delta_{jk} \bi{e}^{\beta}_{il}$. Here,
$\bi{e}^{\beta}_{ij}$ is a basis vector transforming as the index $j$ of the
irreducible representation $\beta$, and $1 \le i \le n^{\beta}_{D}$,
where $n^{\beta}_{D}$ is the multiplicity of $\beta$ in $D$.  To
split up the full vector space into subspaces spanned by orthogonal irreducible basis
vectors, the vectors $\bi{e}^{\alpha}_{i1} \in \, \mathrm{Im} \,
\bi{P}^{\alpha}_{11}$, $\bi{e}^{\alpha}_{i1} \perp \bi{e}^{\alpha}_{j1}$ for
$i \neq j$, must be calculated. The other basis vectors are then
$\bi{e}^{\alpha}_{ij} = \bi{P}^{\alpha}_{j1} \bi{e}^{\alpha}_{i1}$, where $i
\in \{ 1,\ldots,n^{\alpha}_{D} \}$, $j \in \{ 2, \ldots d_{\alpha} \}$. 
The operators $\bi{D}(g)$ are calculated as orthogonal transformation 
matrices for the components of Table \ref{tabone}.

Herewith, the third-order invariants can be found
directly from the full spaces of Table \ref{tabone}. However, we picked the third-order 
invariants by the following procedure: Calculate the second-order irreducible 
components for the irreducible representations $\Gamma^{1}$, $\Gamma^{4}$ and
$\Gamma^{5}$ occuring in the first and third series of (\ref{eq11}). Then set up
all possible invariant scalar products with irreducible components of 
$\bfeta^{u}$ and $\bfeta^{w}$ and throw away occuring linearly dependent 
invariants. 

Since the components for different irreducible representations
do not mix under transformation, another very elegant method,
leading immediately to the ordering with respect to irreducible 
representations described below, is to expand the triple
products of (\ref{eq11}) completely, for example
$[(\Gamma^{4} \oplus \Gamma^{5})\otimes
(\Gamma^{4} \oplus \Gamma^{5})\otimes(\Gamma^{4} \oplus 
\Gamma^{5})]_{s} = (\Gamma^{4} \otimes \Gamma^{4} \otimes \Gamma^{4})_{s} \oplus
[(\Gamma^{4} \otimes \Gamma^{4})_{s} \otimes \Gamma^{5}] \oplus
[\Gamma^{4} \otimes (\Gamma^{5} \otimes \Gamma^{5})_{s}] \oplus
(\Gamma^{5} \otimes \Gamma^{5} \otimes \Gamma^{5})_{s} \,$, and search the
invariants for each of the arising triple products of irreducible
representations separately. 
 
The third-order invariants are listed in \ref{A2}. They consist of components 
of as few different irreducible representations as possible 
(see Table \ref{tabthree}). For each of the four cases, they are
orthonormal, i.e. $\sum_{k=1}^{dim} v_{i,k} v_{j,k} = \delta_{ij}$, where
$dim$ are the respective numbers in the second column of Table \ref{tabone} and
$v_{i,k}$ is the coefficient of the third-order term $k$ in the invariant
$i$. Furthermore, we have tried on the one
hand to choose the invariants as short as possible and on the other
hand to bring them to a suitable form for comparing with each
other and with the invariants of \cite{J14} 
(see \ref{A2} and \ref{A3} for details).

\section{Discussion}

From (\ref{eq11}), there are 20 independent third-order elastic
invariants, or elastic constants, describing the non-linear elasticity
of icosahedral QC.  Since there exist four $uuw$-invariants, we have four 
non-linear phonon-phason couplings. The other third-order invariants are
unsuitable if one wants to generate phason strains or stresses. Despite of
this, from all the third-order invariants, the four $uuu$ ones will play the
most important role because of their influence on the phonon wave 
propagation. To our knowledge, these third-order phonon elastic constants 
have not been determined for QC so far.

Generating phason $\Gamma^{4}$-strains from
phonon stresses now is possible through the invariant $I^{uuw}_{1}$.
Due to (\ref{eq8}),
\begin{equation}
\label{eq12}
\eqalign{
  \eta^{w}_{1} = & \,
  \frac{S^{uuw}_{1}}{20} \sqrt {30} \, \big[3 \, 
  (t_{2}^{u})^{2}+ 3 \, (t_{3}^{u})^{2}- 2\, (t_{4}^{u})^{2} 
  - 2\, (t_{5}^{u})^{2} -2\,(t_{6}^{u})^{2} \big] , 
  \\
  \eta^{w}_{2} = & \,
  -\frac{S^{uuw}_{1}}{2} \sqrt{3} \, \big( 2 \, t_{3}^{u}
  t_{4}^{u} + \sqrt{2} \, t_{5}^{u} t_{6}^{u} \big) , 
  \\
  \eta^{w}_{3} = & \,
  \frac{S^{uuw}_{1}}{2} \, \big( 3 \, t_{2}^{u} t_{5}^{u} 
  + \sqrt{3} \, t_{3}^{u} t_{5}^{u} 
  - \sqrt{6} \, t_{4}^{u} t_{6}^{u} \big) , 
  \\
  \eta^{w}_{4} = & \,
  -\frac{S^{uuw}_{1}}{2} \, \big( 3 \, t_{2}^{u} t_{6}^{u} -
  \sqrt{3} \, t_{3}^{u} t_{6}^{u} + \sqrt{6} \,
  t_{4}^{u} t_{5}^{u} \big) .}
\end{equation}
Here, a perhaps unexpected factor of 3 is present because of the summation rule
in (\ref{eq8}). From (\ref{eq12}), it is obvious that phason 
$\Gamma^{4}$-strains arise from shear stresses $t^{u}_{2}, \ldots, t^{u}_{6}$.
$\eta^{w}_{1}$ is obtained, for example, by applying the stress $t^{u}_{3}
\equiv t$ and all other $t^{u}_{j}=0$, $\eta^{w}_{2}$ by applying
$t^{u}_{3} \equiv t$, $t^{u}_{4} \equiv \sqrt{3/2} \, t$,
$\eta^{w}_{3}$ by applying $t^{u}_{3} \equiv t$, $t^{u}_{5} \equiv
\sqrt{3/2} \, t$ and $\eta^{w}_{4}$ via $t^{u}_{3} \equiv t$,
$t^{u}_{6} \equiv \sqrt{3/2} \, t$. The magnitude of an eventually existing
phonon $\Gamma^{1}$-stress, which is hydrostatic pressure $t^{u}_{1}$, 
has only an indirect effect on the phason $\Gamma^{4}$ strains, due to 
the pressure dependence of elastic compliances. Note that the 
$\Gamma^{4}$-symmetry is unlikely to exist without
simultaneous phason $\Gamma^{5}$-symmetry, which is generated by shear
stresses even in the linear, but also in the non-linear regime according to
higher order compliances (see Table \ref{tabthree}).

Another possibility to obtain phason $\Gamma^{4}$-strains is, 
for example, the quartic electrostriction \cite{J22}.

\appendix

\section{Icosahedral irreducible strains}
\label{A1}
\setcounter{section}{1}

The icosahedral irreducible strain components given below are from
\cite{J21}. They refer to the same coordinate systems as in
\cite{J15,J19,J22}. In Table \ref{tabtwo}, the icosahedral irreducible 
transformation matrices are given, which the irreducible
strains and stresses transform after. Other coordinate systems in use 
are compared to our in some detail in \cite{J15}.
\begin{eqnarray}
\eqalign{
  \bfeta^{u,1} & = \eta^{u}_{1} = \frac{1}{\sqrt{3}} \,
  (\eta^{u}_{11} + \eta^{u}_{22} +
  \eta^{u}_{33})  \, , \\
  \bfeta^{u,5} & = 
\left(
  \begin{array}{c}
  \eta^{u}_{2} \\
  \eta^{u}_{3} \\
  \eta^{u}_{4} \\
  \eta^{u}_{5} \\
  \eta^{u}_{6}
  \end{array} 
\right) =
\left(
  \begin{array}{c}
  \frac{1}{2\sqrt{3}} \, (- \tau^{2} \eta^{u}_{11} + 
  \frac{1}{\tau^{2}} \eta^{u}_{22} +
  (\tau+\frac{1}{\tau}) \eta^{u}_{33}) \\
  \frac{1}{2} ( \frac{1}{\tau} \eta^{u}_{11} - 
  \tau \eta^{u}_{22} +
  \eta^{u}_{33}) \\
  \frac{1}{\sqrt{2}} \, (\eta^{u}_{12}+\eta^{u}_{21}) \\
  \frac{1}{\sqrt{2}} \, (\eta^{u}_{23}+\eta^{u}_{32}) \\
  \frac{1}{\sqrt{2}} \, (\eta^{u}_{31}+\eta^{u}_{13})
  \end{array}
\right) , \\
\bfeta^{w,4} & = \left(
  \begin{array}{c}
  \eta^{w}_{1} \\
  \eta^{w}_{2} \\
  \eta^{w}_{3} \\
  \eta^{w}_{4}
  \end{array} 
\right) = \frac{1}{\sqrt{3}} \left(
  \begin{array}{c}
  \eta^{w}_{11} + 
  \eta^{w}_{22} + \eta^{w}_{33} \\
  \frac{1}{\tau} \eta^{w}_{21} + \tau \eta^{w}_{12} \\
  \frac{1}{\tau} \eta^{w}_{32} + \tau \eta^{w}_{23} \\
  \frac{1}{\tau} \eta^{w}_{13} + \tau \eta^{w}_{31} 
  \end{array}
\right) , \\
  \bfeta^{w,5} &=
\left(
  \begin{array}{c}
  \eta^{w}_{5} \\
  \eta^{w}_{6} \\
  \eta^{w}_{7} \\
  \eta^{w}_{8} \\
  \eta^{w}_{9}
  \end{array} 
\right)
   = \frac{1}{\sqrt{6}} \left(
  \begin{array}{c}
  \sqrt{3} \, ( \eta^{w}_{11} - 
  \eta^{w}_{22} ) \\
  \eta^{w}_{11} + 
  \eta^{w}_{22} - 2 \, \eta^{w}_{33} \\
  \sqrt{2} \, ( \tau \eta^{w}_{21} - 
  \frac{1}{\tau} \eta^{w}_{12}) \\
  \sqrt{2} \, ( \tau \eta^{w}_{32} - 
  \frac{1}{\tau} \eta^{w}_{23} ) \\
  \sqrt{2} \, ( \tau \eta^{w}_{13} - 
  \frac{1}{\tau} \eta^{w}_{31}) 
  \end{array}
\right) .}
\end{eqnarray}

\begin{table}
\fl \caption{Tabulation of the transformation matrices for the
icosahedral irreducible representations. They are given for two
appropriate generating elements of $Y$, i.e. a fivefold rotation
$C_{5}$ and a threefold $C_{3}$. $\tau -1 = \frac{1}{\tau}$. For $Y_{h}$, 
the inversion $i$ must be included. $i$ doesn't change 
the strains \cite{J22}.
\label{tabtwo}} 
\begin{indented}
\item[]
\fl
\begin{tabular}{@{}ccc}
\br                              
$\Gamma$ & $g=C_{5}$ & $g=C_{3}$ \\
\mr
$\Gamma^{1} \equiv 1$ & 1 & 1 \\ \ms
$\Gamma^{3} \equiv 3$ 
& $ \frac{1}{2} \left(
\begin{array}{ccc} 
\tau & \tau-1 & -1 \\ 
\tau-1 & 1 & \tau \\
1 & -\tau & \tau-1 
\end{array} \right) $
& $ \left(
\begin{array}{ccc} 
0 & 0 & 1 \\ 
1 & 0 & 0 \\ 
0 & 1 & 0 
\end{array} \right) $ 
\\ \ms
$\Gamma^{3'} \equiv 3'$ 
& $ \frac{1}{2} \left(
\begin{array}{ccc}
1-\tau & -\tau & -1 \\ 
-\tau & 1 & 1-\tau \\
1 & \tau-1 & -\tau 
\end{array} \right) $  
& $ \left(
\begin{array}{ccc} 
0 & 0 & 1 \\
1 & 0 & 0 \\ 
0 & 1 & 0 
\end{array} \right) $
\\ \ms
$\Gamma^{4} \equiv 4$ 
& $ \frac{1}{4}\left(
\begin{array}{cccc} 
-1 & -\sqrt{5} & \sqrt{5} & -\sqrt{5} \\ 
-\sqrt{5} & -1 & -3 & -1 \\ 
-\sqrt{5} & 3 & 1 & -1 \\
\sqrt{5} & 1 & -1 & -3  
\end{array} \right) $
& $ \left(
\begin{array}{cccc} 
1 & 0 & 0 & 0 \\ 
0 & 0 & 0 & 1 \\ 
0 & 1 & 0 & 0 \\
0 & 0 & 1 & 0  
\end{array} \right) $
\\ \ms
$\Gamma^{5}  \equiv 5$ 
& $ \frac{1}{4} \left(
\begin{array}{ccccc} 
1 & -\sqrt{3} & -\sqrt{6} & 0 & \sqrt{6} \\ 
-\sqrt{3} & -1 & -\sqrt{2} & -2\sqrt{2} & -\sqrt{2} \\ 
-\sqrt{6} & -\sqrt{2} & 2 & 0 & 2 \\
0 & 2\sqrt{2} & 0 & -2 & 2 \\
-\sqrt{6} & \sqrt{2} & -2 & 2 & 0  
\end{array} \right) $
& $ \frac{1}{2} \left(
\begin{array}{ccccc} 
-1 & -\sqrt{3} & 0 & 0 & 0 \\ 
\sqrt{3} & -1 & 0 & 0 & 0 \\ 
0 & 0 & 0 & 0 & 2 \\
0 & 0 & 2 & 0 & 0 \\
0 & 0 & 0 & 2 & 0  
\end{array} \right) $ \\
\br
\end{tabular}
\end{indented}
\end{table}
\section{Third-order icosahedral invariants}
\label{A2}
\setcounter{section}{2}
\begin{eqnarray}
\fl I^{uuu}_{1}= (\eta^{u}_{1})^{3},
\nonumber \\ \ms
\fl I^{uuu}_{2}= \frac{1}{20}\sqrt{30}\,\big[
-(\eta^{u}_{3})^{3}
+\eta^{u}_{3}(\eta^{u}_{5})^{2}
+\eta^{u}_{3}(\eta^{u}_{6})^{2}
-2\,\eta^{u}_{3}(\eta^{u}_{4})^{2}
+3\,\eta^{u}_{3}(\eta^{u}_{2})^{2}
+4\sqrt{2}\,\eta^{u}_{4}\eta^{u}_{5}\eta^{u}_{6}
\nonumber \\ \ns\ns
+\sqrt{3}\,\eta^{u}_{2}(\eta^{u}_{5})^{2}
-\sqrt{3}\,\eta^{u}_{2}(\eta^{u}_{6})^{2}
\big],
\nonumber \\ 
\fl I^{uuu}_{3}=\frac{1}{20}\sqrt{10}\,\big[(\eta^{u}_{2})^{3}
-3\,\eta^{u}_{2}(\eta^{u}_{3})^{2}
-3\,\eta^{u}_{2}(\eta^{u}_{5})^{2}
-3\,\eta^{u}_{2}(\eta^{u}_{6})^{2}
+6\,\eta^{u}_{2}(\eta^{u}_{4})^{2}
+3\sqrt{3}\,\eta^{u}_{3}(\eta^{u}_{5})^{2}
\nonumber \\ \ns\ns
-3\sqrt{3}\,\eta^{u}_{3}(\eta^{u}_{6})^{2}\big], 
\nonumber \\ \ms 
\fl I^{uuu}_{4}=\frac{1}{5}\sqrt{15}\,\eta^{u}_{1}
\big[(\eta^{u}_{2})^{2}+(\eta^{u}_{3})^{2}+
(\eta^{u}_{4})^{2}+(\eta^{u}_{5})^{2}+
(\eta^{u}_{6})^{2}\big],
\nonumber \\
\\
\fl I^{uuw}_{1}=\frac{1}{60}\sqrt{30}\,\big[
2\,(\eta^{u}_{4})^{2}\eta^{w}_{1}
+2\,(\eta^{u}_{5})^{2}\eta^{w}_{1}
+2\,(\eta^{u}_{6})^{2}\eta^{w}_{1}
-3\,(\eta^{u}_{2})^{2}\eta^{w}_{1}
-3\,(\eta^{u}_{3})^{2}\eta^{w}_{1}
\nonumber\\\ns\ns
+2\sqrt{5}\,\eta^{u}_{5}\eta^{u}_{6}\eta^{w}_{2}
+2\sqrt{5}\,\eta^{u}_{4}\eta^{u}_{6}\eta^{w}_{3}
+2\sqrt{5}\,\eta^{u}_{4}\eta^{u}_{5}\eta^{w}_{4}
-\sqrt{10}\,\eta^{u}_{3}\eta^{u}_{5}\eta^{w}_{3}
\nonumber\\
-\sqrt{10}\,\eta^{u}_{3}\eta^{u}_{6}\eta^{w}_{4}
+2\sqrt{10}\,\eta^{u}_{3}\eta^{u}_{4}\eta^{w}_{2}
-\sqrt{30}\,\eta^{u}_{2}\eta^{u}_{5}\eta^{w}_{3}
+\sqrt{30}\,\eta^{u}_{2}\eta^{u}_{6}\eta^{w}_{4}
\big],
\nonumber \\\ms
\fl I^{uuw}_{2}=\frac{1}{60}\sqrt{30}\,\big[
(\eta^{u}_{5})^{2}\eta^{w}_{6}
+(\eta^{u}_{6})^{2}\eta^{w}_{6}
-2\,(\eta^{u}_{4})^{2}\eta^{w}_{6}
+2\,\eta^{u}_{3}\eta^{u}_{5}\eta^{w}_{8}
+2\,\eta^{u}_{3}\eta^{u}_{6}\eta^{w}_{9}
+3\,(\eta^{u}_{2})^{2}\eta^{w}_{6}
\nonumber \\ \ns\ns
-3\,(\eta^{u}_{3})^{2}\eta^{w}_{6}
-4\,\eta^{u}_{3}\eta^{u}_{4}\eta^{w}_{7}
+6\,\eta^{u}_{2}\eta^{u}_{3}\eta^{w}_{5}
+4\sqrt{2}\,\eta^{u}_{5}\eta^{u}_{6}\eta^{w}_{7}
+4\sqrt{2}\,\eta^{u}_{4}\eta^{u}_{6}\eta^{w}_{8}
\nonumber\\
+4\sqrt{2}\,\eta^{u}_{4}\eta^{u}_{5}\eta^{w}_{9}
+\sqrt{3}\,(\eta^{u}_{5})^{2}\eta^{w}_{5}
-\sqrt{3}\,(\eta^{u}_{6})^{2}\eta^{w}_{5}
+2\sqrt{3}\,\eta^{u}_{2}\eta^{u}_{5}\eta^{w}_{8}
\nonumber\\
-2\sqrt{3}\,\eta^{u}_{2}\eta^{u}_{6}\eta^{w}_{9}
\big],
\nonumber \\
\fl I^{uuw}_{3}=\frac{1}{20}\sqrt{10}\,
\big[(\eta^{u}_{2})^{2}\eta^{w}_{5}
-(\eta^{u}_{3})^{2}\eta^{w}_{5}
-(\eta^{u}_{5})^{2}\eta^{w}_{5}
-(\eta^{u}_{6})^{2}\eta^{w}_{5}
+2\,(\eta^{u}_{4})^{2}\eta^{w}_{5}
-2\,\eta^{u}_{2}\eta^{u}_{3}\eta^{w}_{6}
\nonumber \\\ns\ns
-2\,\eta^{u}_{2}\eta^{u}_{5}\eta^{w}_{8}
-2\,\eta^{u}_{2}\eta^{u}_{6}\eta^{w}_{9}
+4\,\eta^{u}_{2}\eta^{u}_{4}\eta^{w}_{7}
+\sqrt{3}\,(\eta^{u}_{5})^{2}\eta^{w}_{6}
-\sqrt{3}\,(\eta^{u}_{6})^{2}\eta^{w}_{6}
\nonumber \\
+2\sqrt{3}\,\eta^{u}_{3}\eta^{u}_{5}\eta^{w}_{8}
-2\sqrt{3}\,\eta^{u}_{3}\eta^{u}_{6}\eta^{w}_{9}
\big],
\nonumber \\\ms
\fl I^{uuw}_{4}=\frac{1}{5}\sqrt{10}\,\eta^{u}_{1}\big(
\eta^{u}_{2}\eta^{w}_{5}+
\eta^{u}_{3}\eta^{w}_{6}+\eta^{u}_{4}\eta^{w}_{7}
+\eta^{u}_{5}\eta^{w}_{8}+
\eta^{u}_{6}\eta^{w}_{9}\big), 
\nonumber \\ 
\\ 
\fl I^{uww}_{1}=\frac{1}{2}\, \eta^{u}_{1}\big[
(\eta^{w}_{1})^{2}+(\eta^{w}_{2})^{2}+(\eta^{w}_{3})^{2}
+(\eta^{w}_{4})^{2}
\big],
\nonumber \\\ms
\fl I^{uww}_{2}=\frac{1}{5}\sqrt{5}\,\eta^{u}_{1}
\big[(\eta^{w}_{5})^{2}+(\eta^{w}_{6})^{2}+
(\eta^{w}_{7})^{2}+(\eta^{w}_{8})^{2}+
(\eta^{w}_{9})^{2}\big],
\nonumber \\\ms
\fl I^{uww}_{3}=\frac{1}{60}\sqrt{30}\,\big[
\eta^{u}_{3}(\eta^{w}_{8})^{2}
+\eta^{u}_{3}(\eta^{w}_{9})^{2}
-2\,\eta^{u}_{3}(\eta^{w}_{7})^{2}
+2\,\eta^{u}_{5}\eta^{w}_{6}\eta^{w}_{8}
+2\,\eta^{u}_{6}\eta^{w}_{6}\eta^{w}_{9}
+3\,\eta^{u}_{3}(\eta^{w}_{5})^{2}
\nonumber \\\ns\ns
-3\,\eta^{u}_{3}(\eta^{w}_{6})^{2}
-4\,\eta^{u}_{4}\eta^{w}_{6}\eta^{w}_{7}
+6\,\eta^{u}_{2}\eta^{w}_{5}\eta^{w}_{6}
+4\sqrt{2}\,\eta^{u}_{5}\eta^{w}_{7}\eta^{w}_{9}
+4\sqrt{2}\,\eta^{u}_{6}\eta^{w}_{7}\eta^{w}_{8}
\nonumber \\
+4\sqrt{2}\,\eta^{u}_{4}\eta^{w}_{8}\eta^{w}_{9}
-\sqrt{3}\,\eta^{u}_{2}(\eta^{w}_{9})^{2}
+\sqrt{3}\,\eta^{u}_{2}(\eta^{w}_{8})^{2}
-2\sqrt{3}\,\eta^{u}_{6}\eta^{w}_{5}\eta^{w}_{9}
\nonumber \\
+2\sqrt{3}\,\eta^{u}_{5}\eta^{w}_{5}\eta^{w}_{8}\big],
\nonumber \\\ms
\fl I^{uww}_{4}=\frac{1}{20}\sqrt{10}\,\big[
\eta^{u}_{2}(\eta^{w}_{5})^{2}-\eta^{u}_{2}(\eta^{w}_{6})^{2}
-\eta^{u}_{2}(\eta^{w}_{8})^{2}
-\eta^{u}_{2}(\eta^{w}_{9})^{2}
+2\,\eta^{u}_{2}(\eta^{w}_{7})^{2}
-2\,\eta^{u}_{3}\eta^{w}_{5}\eta^{w}_{6}
\nonumber\\ \ns\ns
-2\,\eta^{u}_{5}\eta^{w}_{5}\eta^{w}_{8}
-2\,\eta^{u}_{6}\eta^{w}_{5}\eta^{w}_{9}
+4\,\eta^{u}_{4}\eta^{w}_{5}\eta^{w}_{7}
+\sqrt{3}\,\eta^{u}_{3}(\eta^{w}_{8})^{2}
-\sqrt{3}\,\eta^{u}_{3}(\eta^{w}_{9})^{2}
\nonumber\\
+2\sqrt{3}\,\eta^{u}_{5}\eta^{w}_{6}\eta^{w}_{8}
-2\sqrt{3}\,\eta^{u}_{6}\eta^{w}_{6}\eta^{w}_{9}
\big],
\nonumber \\ \ms
\fl I^{uww}_{5}=\frac{1}{30}\sqrt{15}\,\big[
2\,\eta^{u}_{6}\eta^{w}_{2}\eta^{w}_{3}
+2\,\eta^{u}_{5}\eta^{w}_{2}\eta^{w}_{4}
+2\,\eta^{u}_{4}\eta^{w}_{3}\eta^{w}_{4}
+\sqrt{2}\,\eta^{u}_{3}(\eta^{w}_{3})^{2}
+\sqrt{2}\,\eta^{u}_{3}(\eta^{w}_{4})^{2}
\nonumber\\\ns\ns
-2\sqrt{2}\,\eta^{u}_{3}(\eta^{w}_{2})^{2}
+2\sqrt{5}\,\eta^{u}_{4}\eta^{w}_{1}\eta^{w}_{2}
+2\sqrt{5}\,\eta^{u}_{5}\eta^{w}_{1}\eta^{w}_{3}
+2\sqrt{5}\,\eta^{u}_{6}\eta^{w}_{1}\eta^{w}_{4}
\nonumber\\
+\sqrt{6}\,\eta^{u}_{2}(\eta^{w}_{3})^{2}
-\sqrt{6}\,\eta^{u}_{2}(\eta^{w}_{4})^{2}
\big],
\nonumber \\\ms
\fl I^{uww}_{6}=\frac{1}{60}\sqrt{15}\,\big(
4\,\eta^{u}_{4}\eta^{w}_{1}\eta^{w}_{7}
+4\,\eta^{u}_{5}\eta^{w}_{1}\eta^{w}_{8}
+4\,\eta^{u}_{6}\eta^{w}_{1}\eta^{w}_{9}
-6\,\eta^{u}_{2}\eta^{w}_{1}\eta^{w}_{5}
-6\,\eta^{u}_{3}\eta^{w}_{1}\eta^{w}_{6}
\nonumber \\ \ns\ns
+2\sqrt{5}\,\eta^{u}_{6}\eta^{w}_{2}\eta^{w}_{8}
+2\sqrt{5}\,\eta^{u}_{5}\eta^{w}_{2}\eta^{w}_{9}
+2\sqrt{5}\,\eta^{u}_{6}\eta^{w}_{3}\eta^{w}_{7}
+2\sqrt{5}\,\eta^{u}_{4}\eta^{w}_{3}\eta^{w}_{9}
\nonumber \\
+2\sqrt{5}\,\eta^{u}_{5}\eta^{w}_{4}\eta^{w}_{7}
+2\sqrt{5}\,\eta^{u}_{4}\eta^{w}_{4}\eta^{w}_{8}
-\sqrt{10}\,\eta^{u}_{5}\eta^{w}_{3}\eta^{w}_{6}
-\sqrt{10}\,\eta^{u}_{3}\eta^{w}_{3}\eta^{w}_{8}
\nonumber\\
-\sqrt{10}\,\eta^{u}_{6}\eta^{w}_{4}\eta^{w}_{6}
-\sqrt{10}\,\eta^{u}_{3}\eta^{w}_{4}\eta^{w}_{9}
+2\sqrt{10}\,\eta^{u}_{4}\eta^{w}_{2}\eta^{w}_{6}
+2\sqrt{10}\,\eta^{u}_{3}\eta^{w}_{2}\eta^{w}_{7}
\nonumber\\
-\sqrt{30}\,\eta^{u}_{5}\eta^{w}_{3}\eta^{w}_{5}
-\sqrt{30}\,\eta^{u}_{2}\eta^{w}_{3}\eta^{w}_{8}
+\sqrt{30}\,\eta^{u}_{6}\eta^{w}_{4}\eta^{w}_{5}
+\sqrt{30}\,\eta^{u}_{2}\eta^{w}_{4}\eta^{w}_{9}\big), 
\nonumber \\\ms
\fl I^{uww}_{7}=\frac{1}{20}\sqrt{10}\,\big(
\eta^{u}_{5}\eta^{w}_{3}\eta^{w}_{5}
-\eta^{u}_{2}\eta^{w}_{3}\eta^{w}_{8}
+\eta^{u}_{6}\eta^{w}_{4}\eta^{w}_{5}
-\eta^{u}_{2}\eta^{w}_{4}\eta^{w}_{9}
-2\,\eta^{u}_{4}\eta^{w}_{2}\eta^{w}_{5}
+2\,\eta^{u}_{2}\eta^{w}_{2}\eta^{w}_{7}
\nonumber\\\ns\ns
-\sqrt{3}\,\eta^{u}_{5}\eta^{w}_{3}\eta^{w}_{6}
+\sqrt{3}\,\eta^{u}_{3}\eta^{w}_{3}\eta^{w}_{8}
+\sqrt{3}\,\eta^{u}_{6}\eta^{w}_{4}\eta^{w}_{6}
-\sqrt{3}\,\eta^{u}_{3}\eta^{w}_{4}\eta^{w}_{9}
\nonumber\\
+\sqrt{6}\,\eta^{u}_{6}\eta^{w}_{2}\eta^{w}_{8}
-\sqrt{6}\,\eta^{u}_{5}\eta^{w}_{2}\eta^{w}_{9}
-\sqrt{6}\,\eta^{u}_{6}\eta^{w}_{3}\eta^{w}_{7}
+\sqrt{6}\,\eta^{u}_{4}\eta^{w}_{3}\eta^{w}_{9}
\nonumber\\
+\sqrt{6}\,\eta^{u}_{5}\eta^{w}_{4}\eta^{w}_{7}
-\sqrt{6}\,\eta^{u}_{4}\eta^{w}_{4}\eta^{w}_{8}
-\sqrt{10}\,\eta^{u}_{3}\eta^{w}_{1}\eta^{w}_{5}
+\sqrt{10}\,\eta^{u}_{2}\eta^{w}_{1}\eta^{w}_{6}\big),
\nonumber \\
\\ 
\fl I^{www}_{1}=\frac{1}{4}\sqrt{3}\,\big[
(\eta^{w}_{1})^{3}
-\eta^{w}_{1}(\eta^{w}_{2})^{2}
-\eta^{w}_{1}(\eta^{w}_{3})^{2}
-\eta^{w}_{1}(\eta^{w}_{4})^{2}+2\sqrt{5}\,
\eta^{w}_{2}\eta^{w}_{3}\eta^{w}_{4}\big],
\nonumber \\\ms
\fl I^{www}_{2}=\frac{1}{20}\sqrt{30}\,\big[
-(\eta^{w}_{6})^{3}
+\eta^{w}_{6}(\eta^{w}_{8})^{2}
+\eta^{w}_{6}(\eta^{w}_{9})^{2}
-2\,\eta^{w}_{6}(\eta^{w}_{7})^{2}
+3\,\eta^{w}_{6}(\eta^{w}_{5})^{2}
\nonumber\\ \ns\ns
+4\sqrt{2}\,\eta^{w}_{7}\eta^{w}_{8}\eta^{w}_{9}
+\sqrt{3}\,\eta^{w}_{5}(\eta^{w}_{8})^{2}
-\sqrt{3}\,\eta^{w}_{5}(\eta^{w}_{9})^{2}
\big],
\nonumber \\\ms
\fl I^{www}_{3}=\frac{1}{20}\sqrt{10}\,\big[(\eta^{w}_{5})^{3}
-3\,\eta^{w}_{5}(\eta^{w}_{6})^{2}
-3\,\eta^{w}_{5}(\eta^{w}_{8})^{2}
-3\,\eta^{w}_{5}(\eta^{w}_{9})^{2}
+6\,\eta^{w}_{5}(\eta^{w}_{7})^{2}
\nonumber \\\ns\ns
+3\sqrt{3}\,\eta^{w}_{6}(\eta^{w}_{8})^{2}
-3\sqrt{3}\,\eta^{w}_{6}(\eta^{w}_{9})^{2}
\big],
\nonumber \\\ms
\fl I^{www}_{4}=\frac{1}{10}\sqrt{10}\,\big[
(\eta^{w}_{3})^{2}\eta^{w}_{6}
+(\eta^{w}_{4})^{2}\eta^{w}_{6}
-2\,(\eta^{w}_{2})^{2}\eta^{w}_{6}
+\sqrt{2}\,\eta^{w}_{2}\eta^{w}_{3}\eta^{w}_{9}
+\sqrt{2}\,\eta^{w}_{2}\eta^{w}_{4}\eta^{w}_{8}
\nonumber\\ \ns\ns
+\sqrt{2}\,\eta^{w}_{3}\eta^{w}_{4}\eta^{w}_{7}
+\sqrt{3}\,(\eta^{w}_{3})^{2}\eta^{w}_{5}
-\sqrt{3}\,(\eta^{w}_{4})^{2}\eta^{w}_{5}
+\sqrt{10}\,\eta^{w}_{1}\eta^{w}_{4}\eta^{w}_{9}
\nonumber \\
+\sqrt{10}\,\eta^{w}_{1}\eta^{w}_{2}\eta^{w}_{7}
+\sqrt{10}\,\eta^{w}_{1}\eta^{w}_{3}\eta^{w}_{8}\big],
\nonumber \\\ms
\fl I^{www}_{5}=\frac{1}{20}\sqrt{10}\,\big[
2\,\eta^{w}_{1}(\eta^{w}_{7})^{2}
+2\,\eta^{w}_{1}(\eta^{w}_{8})^{2}
+2\,\eta^{w}_{1}(\eta^{w}_{9})^{2}
-3\,\eta^{w}_{1}(\eta^{w}_{5})^{2}
-3\,\eta^{w}_{1}(\eta^{w}_{6})^{2}
\nonumber \\ \ns\ns
+2\sqrt{5}\,\eta^{w}_{2}\eta^{w}_{8}\eta^{w}_{9}
+2\sqrt{5}\,\eta^{w}_{3}\eta^{w}_{7}\eta^{w}_{9}
+2\sqrt{5}\,\eta^{w}_{4}\eta^{w}_{7}\eta^{w}_{8}
-\sqrt{10}\,\eta^{w}_{3}\eta^{w}_{6}\eta^{w}_{8}
\nonumber \\
-\sqrt{10}\,\eta^{w}_{4}\eta^{w}_{6}\eta^{w}_{9}
+2\sqrt{10}\,\eta^{w}_{2}\eta^{w}_{6}\eta^{w}_{7}
-\sqrt{30}\,\eta^{w}_{3}\eta^{w}_{5}\eta^{w}_{8}
+\sqrt{30}\,\eta^{w}_{4}\eta^{w}_{5}\eta^{w}_{9}
\big].
\nonumber \\
\end{eqnarray}
The invariants comprise strain components of irreducible
representations as displayed in table B1.
Note that the following pairs of invariants have exactly the same structure:
$I^{uuu}_{2}/I^{www}_{2}$, $I^{uuu}_{3}/I^{www}_{3}$,
$I^{uuu}_{4}/I^{uww}_{2}$, $I^{uuw}_{1}/I^{www}_{5}$,
$I^{uuw}_{2}/I^{uww}_{3}$, $I^{uuw}_{3}/I^{uww}_{4}$,
$I^{uww}_{5}/I^{www}_{4}$. Furthermore, $3\,I^{uww}_{3}$ is obtained from
$I^{www}_{2}$ by replacing all products
$\eta^{w}_{i} \eta^{w}_{j} \eta^{w}_{k}$
by $\eta^{u}_{i-3} \eta^{w}_{j} \eta^{w}_{k} + \eta^{w}_{i}
\eta^{u}_{j-3} \eta^{w}_{k} + \eta^{w}_{i} \eta^{w}_{j}
\eta^{u}_{k-3}$. The same is true for $3\,I^{uww}_{4}$ and
$I^{www}_{3}$. $\sqrt{6}\,I^{uww}_{6}$ follows from $I^{www}_{5}$ by
replacing only the phason $\Gamma^{5}$-components in the manner
described above.

\begin{table}
\caption{Powers of irreducible representation components 
$\eta^{a,\alpha}_{i}$ in third-order invariants. Possible combinations are:
$\alpha = 1$, $i=1$; $\alpha = 5$, $i=2,3,4,5,6$ ($a=u$); 
$\alpha = 4$, $i=1,2,3,4$; $\alpha = 5$, $i=5,6,7,8,9$ ($a=w$).
\label{tabthree}} 
\begin{indented}
  \lineup
\item[]\begin{tabular}{@{}*{10}{c}}
\br                              
$(a,\alpha)$ & (u,1) & (u,5) & (w,4) & (w,5) &  &  (u,1) & (u,5) & (w,4) &
(w,5) \\
\mr
$I^{uuu}_{1}$ & 3 &  &  & & 
$I^{uww}_{3}$ &  & 1 &  & 2 \\
$I^{uuu}_{2}$ &  & 3  &  & & 
$I^{uww}_{4}$ &  & 1 & & 2 \\
$I^{uuu}_{3}$ &  & 3 &  & & 
$I^{uww}_{5}$ &  & 1 & 2 &  \\
$I^{uuu}_{4}$ & 1 & 2 &  & & 
$I^{uww}_{6}$ &  & 1 & 1 & 1 \\
$I^{uuw}_{1}$ &  & 2 & 1 &  & 
$I^{uww}_{7}$ &  & 1 & 1 & 1 \\
$I^{uuw}_{2}$ &  & 2  & &  1 & 
$I^{www}_{1}$ &  &  & 3 & \\
$I^{uuw}_{3}$ &  & 2 &  & 1 & 
$I^{www}_{2}$ &  &  & & 3 \\
$I^{uuw}_{4}$ & 1 & 1 & & 1 & 
$I^{www}_{3}$ &  &  & & 3 \\
$I^{uww}_{1}$ & 1 & & 2 & & 
$I^{www}_{4}$ &  &  & 2 & 1 \\
$I^{uww}_{2}$ & 1 &  &  & 2 & 
$I^{www}_{5}$ &  &  & 1 & 2 \\
\br
\end{tabular}
\end{indented}
\end{table}
\section{Relationship to other third-order invariants and elastic isotropy in
  phonon space}
\label{A3}
\begin{eqnarray}
\label{eqA3}
\eqalign{
\fl  \tr(\bfeta^{u})^{3} = 3 \sqrt{3} \, I^{uuu}_{1} \, , 
  \\
\fl  \tr(\bfeta^{u})\{ \tr[(\bfeta^{u})^{2}] \}_{s}
   = \sqrt{3} \, I^{uuu}_{1} + \sqrt{5} \, I^{uuu}_{4} \, , 
  \\
\fl  \{ \tr [(\bfeta^{u})^{3}]\}_{s}  = \frac{1}{3}\sqrt{3} \,
  I^{uuu}_{1} + \frac{1}{4} \sqrt{30} \, I^{uuu}_{2} -
  \frac{5}{12} \sqrt{6} \, I^{uuu}_{3} + \sqrt{5} \, I^{uuu}_{4} \, , 
  \\
\fl  [\det(\bfeta^{u})]_{s} = \frac{1}{9}\sqrt{3} \, I^{uuu}_{1}
  + \frac{1}{12} \sqrt{30} \, I^{uuu}_{2} - \frac{5}{36} \sqrt{6} 
  \, I^{uuu}_{3} - \frac{1}{6} \sqrt{5} \, I^{uuu}_{4} \, , 
  \\ 
\fl  \det(\bfeta^{w}) =  \frac{4}{9} \, I^{www}_{1} +
  \frac{2}{9} \sqrt{5} \, I^{www}_{2} - \frac{1}{9} \sqrt{15} \,
  I^{www}_{4} + \frac{1}{9} \sqrt{30} \, I^{www}_{5} \, 
. }
\end{eqnarray}
In (\ref{eqA3}), index $s$ denotes that $\eta^{u}_{ij}$ must be replaced by
$\frac{1}{2}(\eta^{u}_{ij} + \eta^{u}_{ji})$. 
Some of our third-order invariants are just proportional to $\eta^{u}_{1}$
times one of the second-order, invariant scalar products given in 
Section \ref{sec3}. Note that $\tr(\bfeta^{w})$ 
is not invariant. Our invariants $I^{uuu}_{1}$ and  $I^{uuu}_{4}$ are also
O(3)-invariants, while $I^{uuu}_{2}$ and  $I^{uuu}_{3}$ must be combined to
$\frac{3}{14}\sqrt{14} \, I^{uuu}_{2} - \frac{1}{14} \sqrt{70} \, I^{uuu}_{3}$ to 
give a linearly independent third O(3)-invariant. In the degenerate case of
phononic isotropy, our elastic constants 
$C^{uuu}_{i}$ can be expressed by sets of classical
third-order elastic constants already in use: 
$C^{uuu}_{1} = \sqrt{3} \, (l +  \frac{1}{9} \, n)$,  
$C^{uuu}_{2} = \frac{1}{12} \sqrt{30} \, n$,
$C^{uuu}_{3} = -\frac{5}{36} \sqrt{6} \, n$,
$C^{uuu}_{4} = \sqrt{5} \, (m - \frac{1}{6} \, n)$ \cite{J9,J13},
$C^{uuu}_{1} =  \sqrt{3} \, (\frac{1}{2} \, \nu_{1} + \nu_{2} +
\frac{4}{9} \, \nu_{3})$,
$C^{uuu}_{2} = \frac{1}{3} \sqrt{30} \, \nu_{3}$,
$C^{uuu}_{3} = -\frac{5}{9} \sqrt{6} \, \nu_{3}$,
$C^{uuu}_{4} = \sqrt{5} \, (\nu_{2} + \frac{4}{3} \, \nu_{3})$ 
\cite{J23}.

Up to normalisation factors, the phason third-order invariants 
$\mathcal{I}'_{5}$, $\mathcal{I}_{5}$,
$\mathcal{J}$ and $\mathcal{J}'$ of Ishii \cite{J14} are transformed
into our $I^{www}_{2}$, $I^{www}_{3}$, $I^{www}_{4}$ and
$I^{www}_{5}$, respectively, by the substitutions $\eta^{w}_{5} \rightarrow
-\frac{1}{4} (\sqrt{10} \, \eta^{w}_{5} + \sqrt{6} \, \eta^{w}_{6})$
and  $\eta^{w}_{6} \rightarrow \frac{1}{4} (-\sqrt{6} \, \eta^{w}_{5} +
\sqrt{10} \, \eta^{w}_{6})$. This is necessary because
in \cite{J14} other irreducible components are used than in \cite{J21}.

\end{document}